\documentclass{scsSimAUDPaperFormat}
\copyrightnotice{
    SimAUD 2020 May 25-27 Online
    
    \copyright\,2020 Society for Modeling \& Simulation International (SCS)
}
\usepackage{balance}        
\usepackage{graphics}        
\usepackage{times}            
\usepackage{url}            
\usepackage{dblfloatfix}    
\usepackage[utf8]{inputenc}
\usepackage{newunicodechar}
\newunicodechar{ℓ}{l}

\makeatletter
\def\url@leostyle{%
  \@ifundefined{selectfont}{\def\UrlFont{\sf}}{\def\UrlFont{\small\bf\ttfamily}}}
\makeatother
\def\pprw{8.5in}
\def\pprh{11in}

\usepackage[pdftex]{hyperref}
\hypersetup{
pdftitle={SCS Conference Proceedings Format},
pdfauthor={LaTeX},
pdfkeywords={SCS, proceedings, archival format},
bookmarksnumbered,
pdfstartview={FitH},
colorlinks,
citecolor=blue,
filecolor=black,
linkcolor=blue,
urlcolor=blue,
breaklinks=true,
}

\newcommand{\affiliation}[1]{\ensuremath{^{\textrm{#1}}}}

\begin{document}
\title{\includegraphics[width=1.0\textwidth]{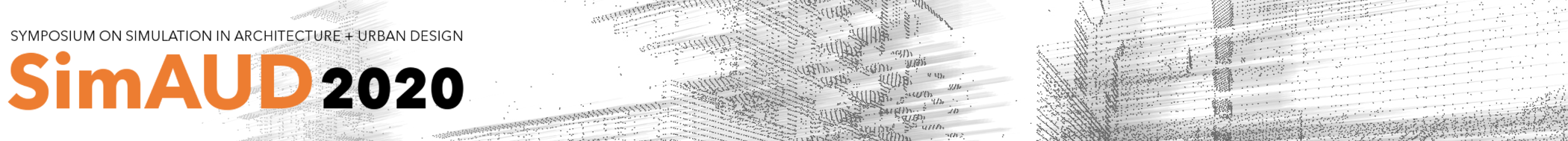}\\\quad\\Build2Vec: Building Representation in Vector Space}
\def\UniversityA{\affiliation{1}} 
\def\CompanyA{\affiliation{2}}
\def\UniversityB{\affiliation{3}}

\author{
}
\author{
	Mahmoud M. Abdelrahman\UniversityA
	,
	Adrian Chong\UniversityA
	,
	Clayton Miller\UniversityA
	\\
	\\
	\affaddr{\UniversityA}{Department of Building, National University of Singapore (NUS), Singapore}}

\maketitle

\begin{abstract}
In this paper, we represent a methodology of a graph embeddings algorithm that is used to transform labeled property graphs obtained from a Building Information Model (BIM). Industrial Foundation Classes (IFC) is a standard schema for BIM, which is utilized to convert the building data into a graph representation. We used node2Vec with biased random walks to extract semantic similarities between different building components and represent them in a multi-dimensional vector space. A case study implementation is conducted on a net-zero-energy building located at the National University of Singapore (SDE4). This approach shows promising machine learning applications in capturing the semantic relations and similarities of different building objects, more specifically, spatial and spatio-temporal data.
\end{abstract}



\section{Introduction}
The amount of data generated during the last two decades exceeds that which has ever been generated in history. This growth is due to the radical evolution of Internet of Things (IoT) networks of interconnected objects that are used in sensing the surrounding environment (sensors) or controlling the physical world (actuators) \cite{Gubbi2013InternetDirections}. Also, more powerful computational power and algorithms have helped in managing and processing these data to extract information \cite{Hilbert2016BigChallenges,Sagiroglu2013BigReview}. In the building industry, IoT, computational resources, and algorithms have pushed our understanding of different scales and interactions of the built environment. Scales, in this context include city, group-of-buildings (district), building, and human scale in addition to the network of interactions between those different scales \cite{Talari2017AConcept}. In this research, we focus on the extraction of spatio-temporal data from buildings and the representation of them in an embedded vector space using graph-embeddings \cite{ZhouGraphApplications,Yang2016RevisitingEmbeddings,ZhangLearningGraphs}. We refer to spatial data such as, spaces, walls, doors and windows; and temporal data as IoT sensors such as, indoor environmental data (temperature, humidity, noise), energy consumption data (equipment load, lighting load, HVAC) and occupants data, which could be considered as a movable sensors \cite{Mahdavi2017AnMonitoring}, (presence, thermal comfort, location over time). This paper aims to introduce \emph{Graph-embeddings} as an effective method to capture the spatial and temporal complexity within the buildings and the representation of them in a single vector space for the aim of improving prediction, classification, and recommendation accuracy. Potential applications are presented in the Discussion Section. 
\begin{figure}[ht]
\centering
\includegraphics[width=0.9\columnwidth]{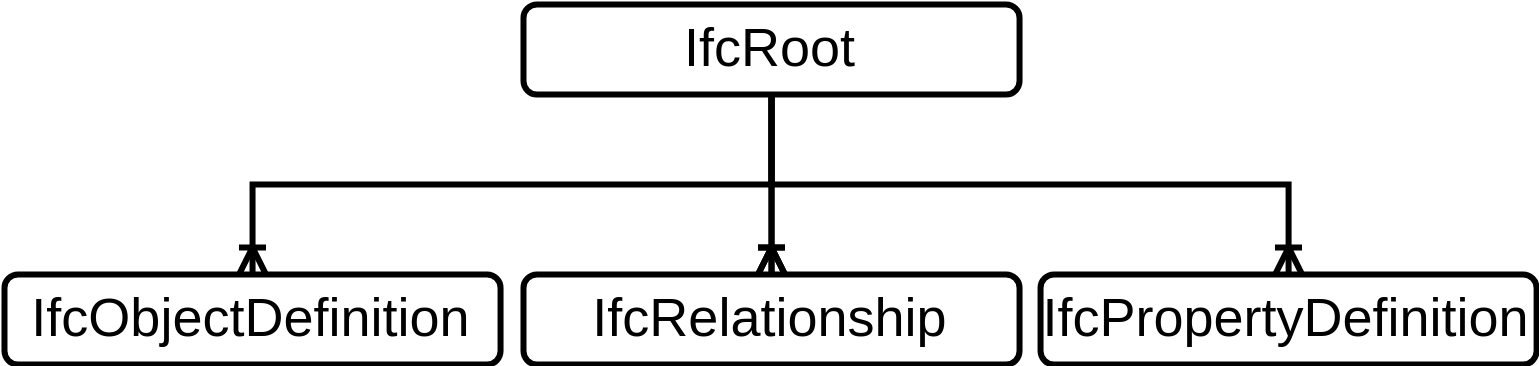}
\caption{Ifc abstract structure}
\label{fig:ifcstructure}
\end{figure}

\begin{figure}[ht]
\centering
\includegraphics[width=0.9\columnwidth]{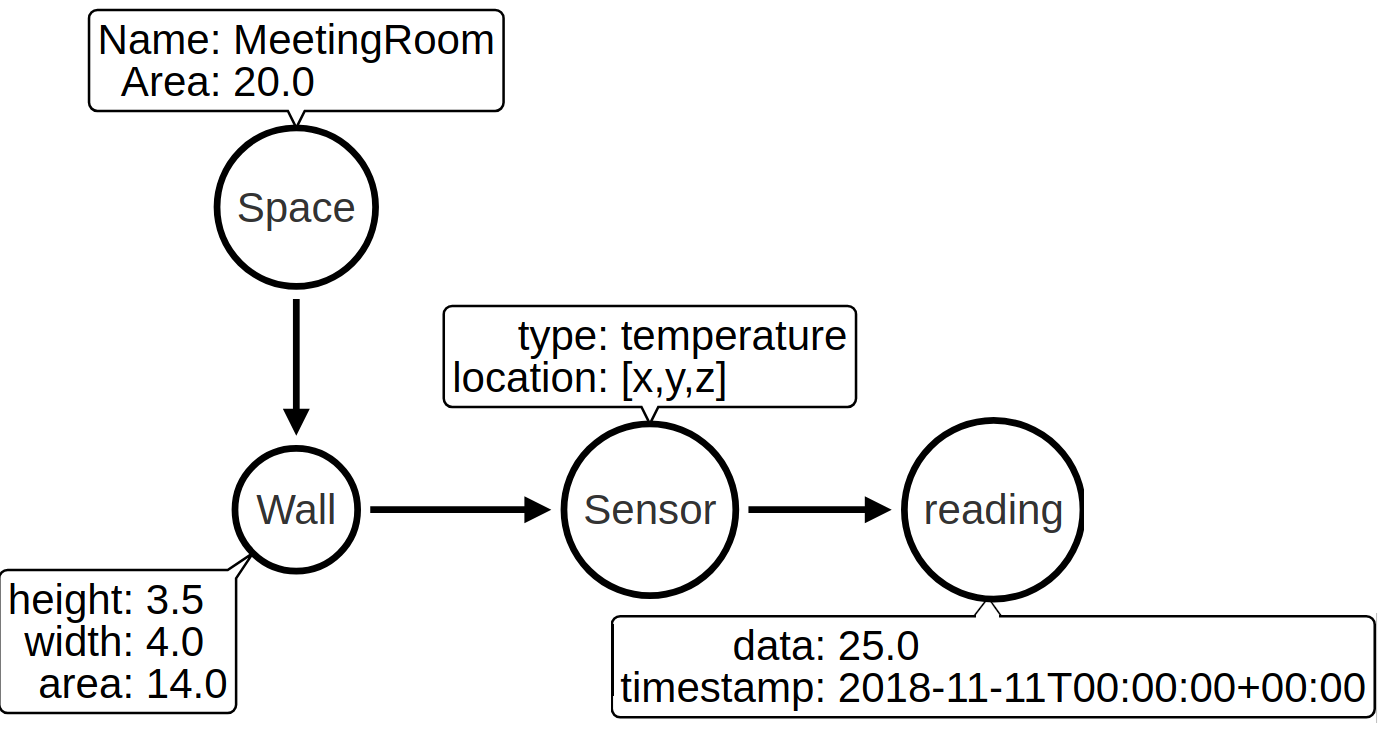}
\caption{Example of attributed graph, where nodes and edges can have additional attributes}
\label{fig:attribgraph}
\end{figure}

\begin{figure*}[ht]
\includegraphics[width=\linewidth]{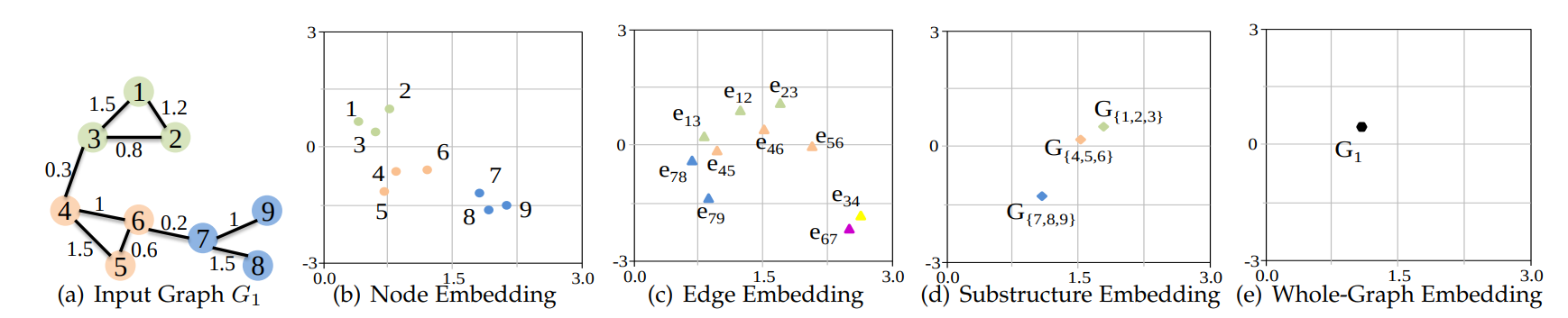}
\caption{Graph embeddings convert the graph structure (vertex and edges) into vector representation. Image adapted from [4]}
\label{fig:node2vec}
\end{figure*}

There are many ways to represent the spatial data of the buildings, amongst which, the Building Information Model (BIM) is the most well-known. BIM models are used in integrated design that enables multiple stakeholders to work on one embedded platform throughout the building life-cycle. This technology enables creating a rich repository of data of the building during its life-cycle, including spatial representations \cite{Eastman2008BIMFabricators, Parn2017TheReview}. BIM models are a static representation of the building; they do not include the temporal data from IoT sensors. Several studies use different methods to fuse the BIM models with IoT data by using middle-ware or representational state transfer (RESTful) API that connects the virtual BIM sensor with the corresponding IoT node \cite{IsikdagBIMPerspective, Bottaccioli2017BuildingModels}.

There are many software/tools available for creating, editing, and managing BIM models. In this study, we use the Industrial Foundation Classes (IFC) which is an open, international standard (ISO 16739-1:2018) digital description of the built environment \cite{IndustryTechnical}. The IFC model consists of objects built in a hierarchical structure, which represents the physical components of the buildings (IFCObjects), the relationships between different objects (IFCRelationship), and the properties of objects (IFCPropertyDefinition) as shown in Figure \ref{fig:ifcstructure}. The hierarchical structure of IFC enables extracting the entity data as a graph. A \emph{graph} is a data structure that consists of a diagrammatic representation or objects and the relation between them in the form of \emph{nodes} (referred to as a vertex) and \emph{edges} connecting those nodes \cite{West1996IntroductionTheory, Gross2004HandbookTheory}. However, a simple graph that consists of vertex and edges is not sufficient to represent the complexity of the BIM data (relations and attributes), and the IoT temporal data. Thus, an attributed graph is more suitable to handle such data. An attributed graph can handle rich information by enabling attributes on both edges and nodes (Figure \ref{fig:attribgraph}). Previous work has introduced methodologies to convert the BIM model to graph theory using an online tool called IFCWebServer \footnote{\href{IFCWebServer.org}{IFCWebServer.org}} \cite{IsmailApplicationStandard, Ismail2018BuildingDatabases}.

Graph embeddings is a method to convert graphs into vectors. An example is seen in Figure \ref{fig:node2vec}. For a weighted graph $G$, where $G(V, E, W)$ is a graph and $V$ denotes vertices, $E$ denotes edges, and $W$ denotes weights: $$GraphEmbedding(G(V,E,W)) \rightarrow {\rm I\!R}^n$$ 
Most machine learning models work with the feature vector representation of data where an instance is multi-dimensional and represents features in different ways (numeric data, binary, categorical,etc.). The motivation behind learning embeddings is to perform important tasks on networks such as classification \cite{Zhang2016LearningTags,ChangHeterogeneousArchitectures,PerozziDeepWalk:Representations}, prediction \cite{GroverNode2vec:Networks} and clustering \cite{Tang2016CappedClustering}. For example, predicting the most probable label of a node in a graph could be used as a recommendation system. Cai et al. conducted an extensive survey on graph embeddings and applications \cite{Cai2017AApplications}. For this research, we will be using two types of graph embeddings, namely, node2Vec \cite{GroverNode2vec:Networks}, and Spatio-Temporal Attentive RNN (STAR) \cite{Xu2019Spatio-TemporalGraphs}. The main difference between these models is that \textbf{node2Vec} works well with static graphs, i.e., not temporal and it does not support directional graphs, while \textbf{STAR} model can capture the temporal attribute changes of the graph which makes it ideal for situations where sensors' temporal data included within the graph network.

\section{Methodology}
\label{sec:methodology}
\subsection{Case study}
Our case-study is the SDE4 building, a net-zero energy building located in the School of Design and Environment (SDE) at the National University of Singapore (NUS). We selected this case study for several reasons:
\begin{enumerate}
    \item Smart building - There are sensors deployed in every space measuring indoor, outdoor environmental data and indoor localization. 
    \item Data Openness: There is to access both spatial data (BIM models) and temporal data (sensors/actuators/weather stations) and thermal comfort data (user feedback).
    \item Variety of HVAC systems: A hybrid cooling system is adopted and it is made up of AC and natural ventilation systems, which give richness to the model.
\end{enumerate}
\subsection{Data sources}
\begin{figure}[ht]
\centering
\includegraphics[width=0.8\columnwidth]{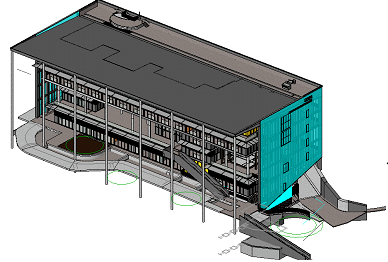}
\caption{The SDE4 case study building BIM model representation}
\label{fig:casestudy}
\end{figure}

There are three main sources of data used in this study: (1) Spatial data: using the IFC file of the BIM model as seen in Figure \ref{fig:casestudy}, (2) Temporal data from the sensors in the form of sequential snapshots, and (3) occupant comfort feedback data from a experimental implementation. 


The IFC file for the SDE4 building is converted from an \emph{Autodesk Revit} model of the building and doesn't include data about the sensors' locations. Additionally, occupants are treated as movable sensors (nodes) whose feedback is the prediction target on hot encoder i.e. comfortable = [1,0,0], uncomfortable=[0,1,0] and neutral=[0,0,1]. These data are not available in the initial version of the IFC file, therefore a preprocessing step of the occupants data should be first performed.

There are two types of sensors deployed into the building spaces: (1) Sensors from the Building Management System (BMS), including in the thermostats that measure the ambient temperature, humidity, and CO2 levels and (2) other non-permanent IoT sensors deployed within the building in different spots that measure air temperature, humidity, light levels, noise levels, and CO2 levels.

The user feedback data were collected during an experiment conducted in the building that included over 30 participants that  gave high-frequency subjective comfort feedback using micro ecological momentary assessments on a smart-watch including their indoor location, and thermal sensation feedback \cite{Jayathissa2020}. The location of each test participant was collected using the YAK mobile indoor localization app \cite{Abdelrahman2019}.

\subsection{Framework implementation example}
\begin{figure}
\centering
\begin{minipage}[b]{.45\columnwidth}
    \centering
\includegraphics[width=\textwidth]{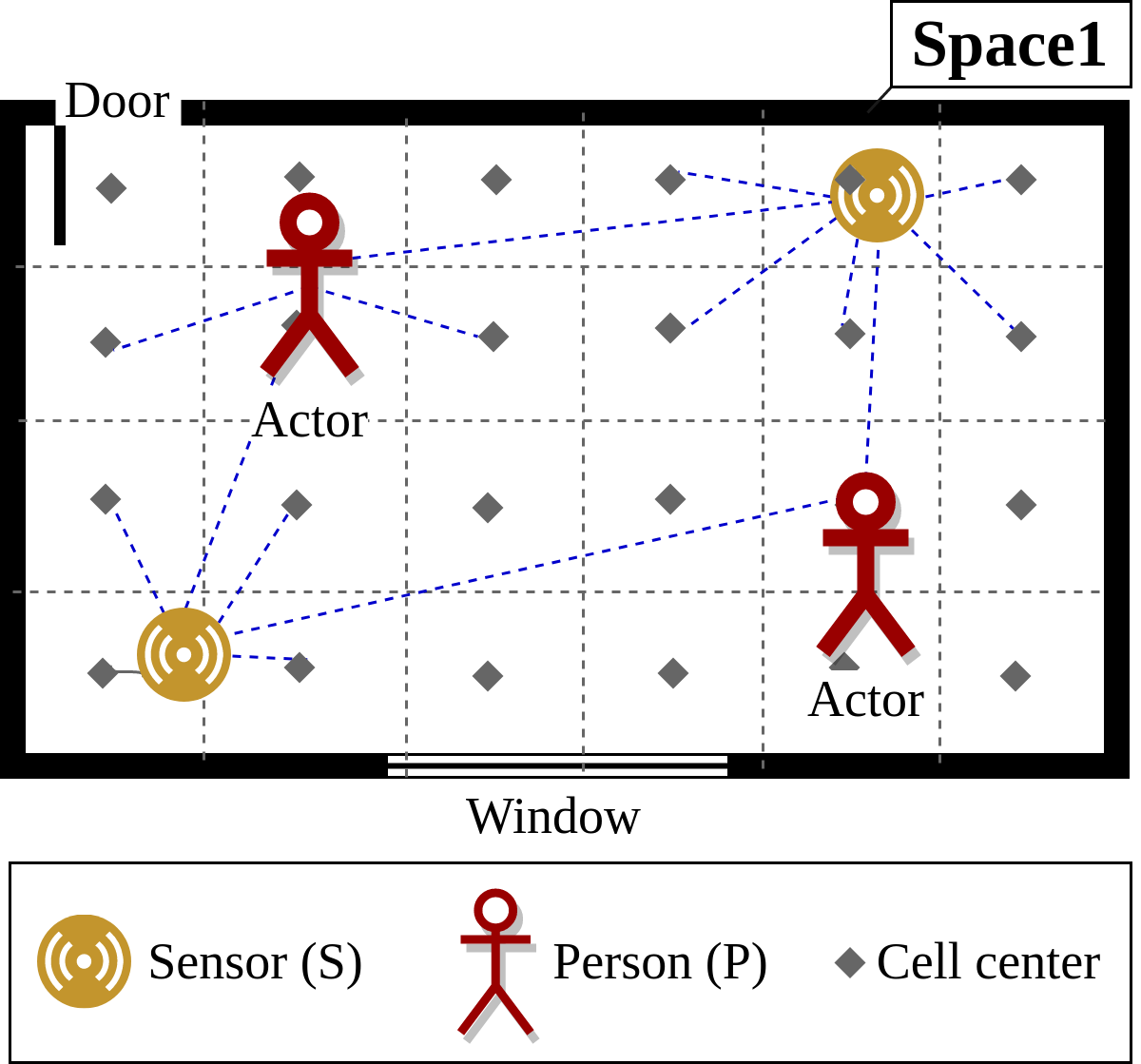}
\caption{Spaces are discretized and objects are dynamically connected to each other.}
\label{fig:ifc2graph}
\end{minipage}\qquad
\begin{minipage}[b]{.45\columnwidth}
\centering
\includegraphics[width=\textwidth]{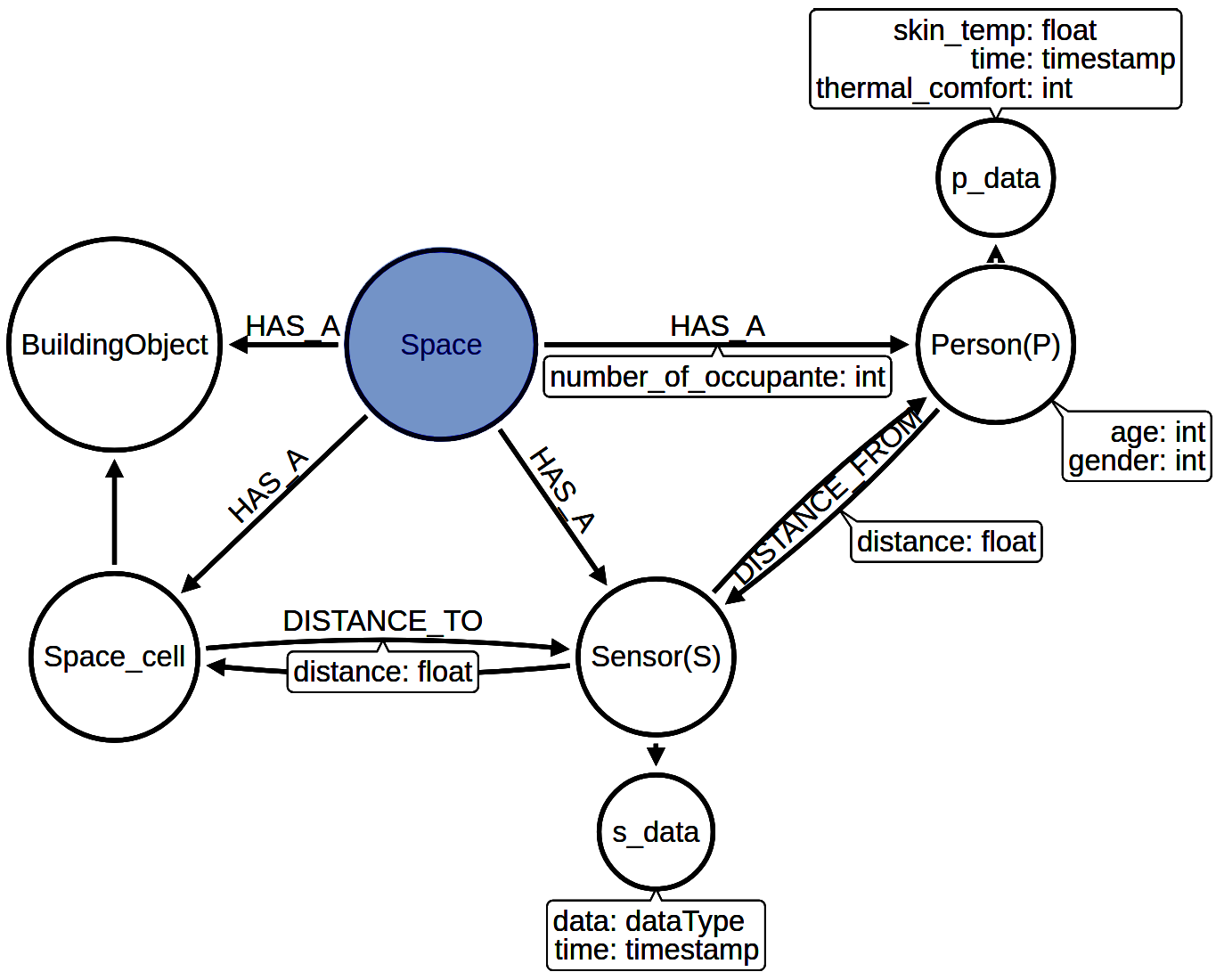}
\caption{A temporal graph is dynamically changing at each time step
\label{fig:framework}}
\end{minipage}
\end{figure}

\begin{figure}[ht]
\centering
\includegraphics[width=0.9\columnwidth]{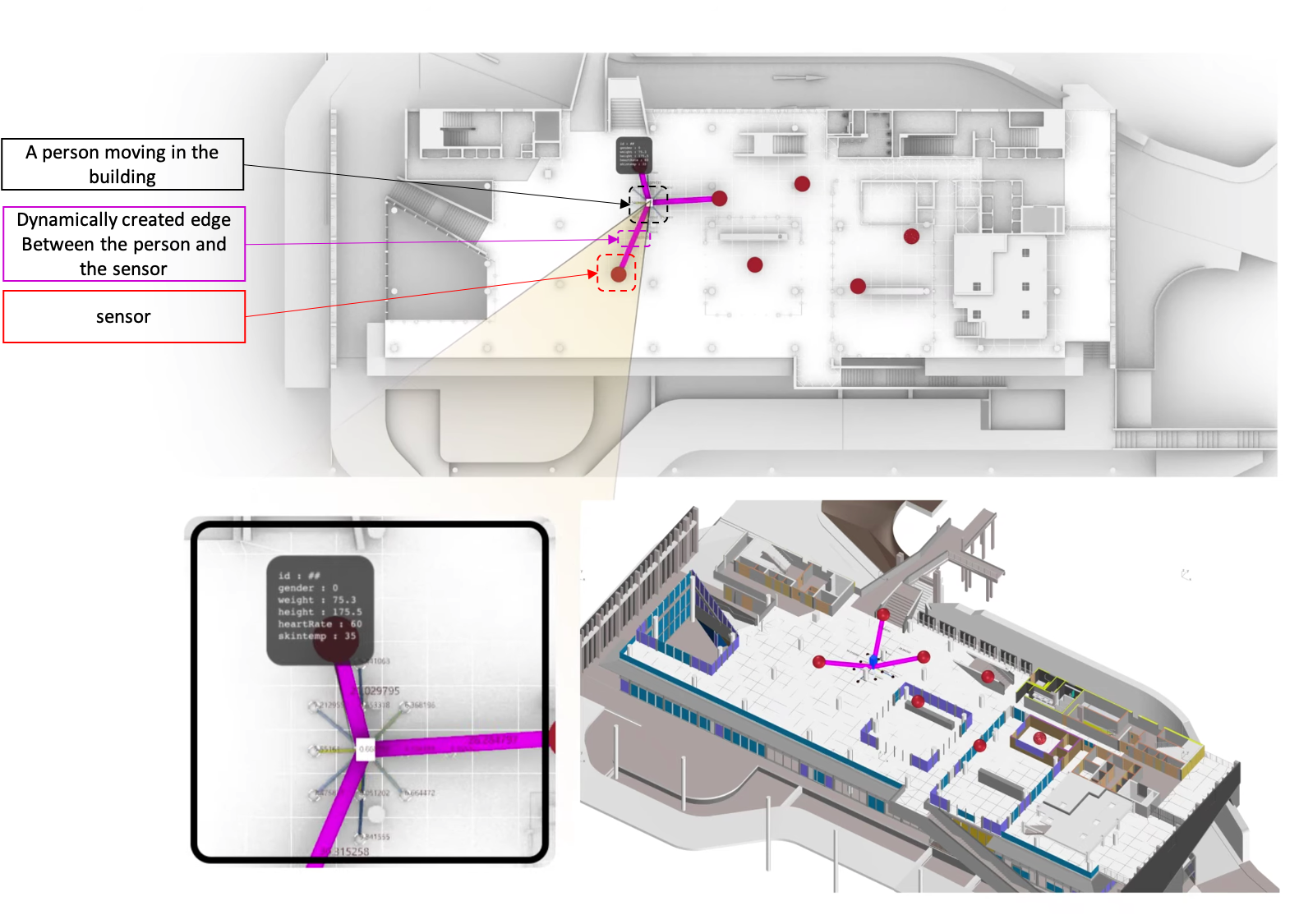}
\caption{Snapshot of the animated demo of the attributed graph schematic design after discretizing the space into cells (link to animated online version found in the text)}
\label{fig:dynamic}
\end{figure}

A sample of the framework is shown in Figures \ref{fig:ifc2graph} and \ref{fig:framework}. If \textit{Space1} from the IFC file is targeted, this space is a supertype of some other objects such as \emph{IfcDoor}, \emph{IfcWindow}, \emph{IfcWalls}, \emph{IfcFloor}, and \emph{IfcCieling}. However, an occupant does not have a corresponding \textit{dynamic} IFC object, so a method was developed to trigger the occupant once they moves into the space. The space is first (discretized) into two-dimensional structured cells using a spatial-discretization algorithm \cite{Fischer1993FiniteAreas}. Spatial discretization divides the space shape into finite elements in the form of graph objects. Secondly, the relation between a moving person in space and the adjacent cells is created at each time-step as shown in the snapshot of Figure \ref{fig:dynamic}. Similarly, a relation (edge) between each sensor and the adjacent cells is created. An animated illustration of Figure \ref{fig:dynamic} can be viewed online\footnote{\href{https://youtu.be/4iFuQvKG\_Wg}{https://youtu.be/4iFuQvKG\_Wg}}. 


\begin{figure}[h!]
\centering
\includegraphics[width=0.9\columnwidth]{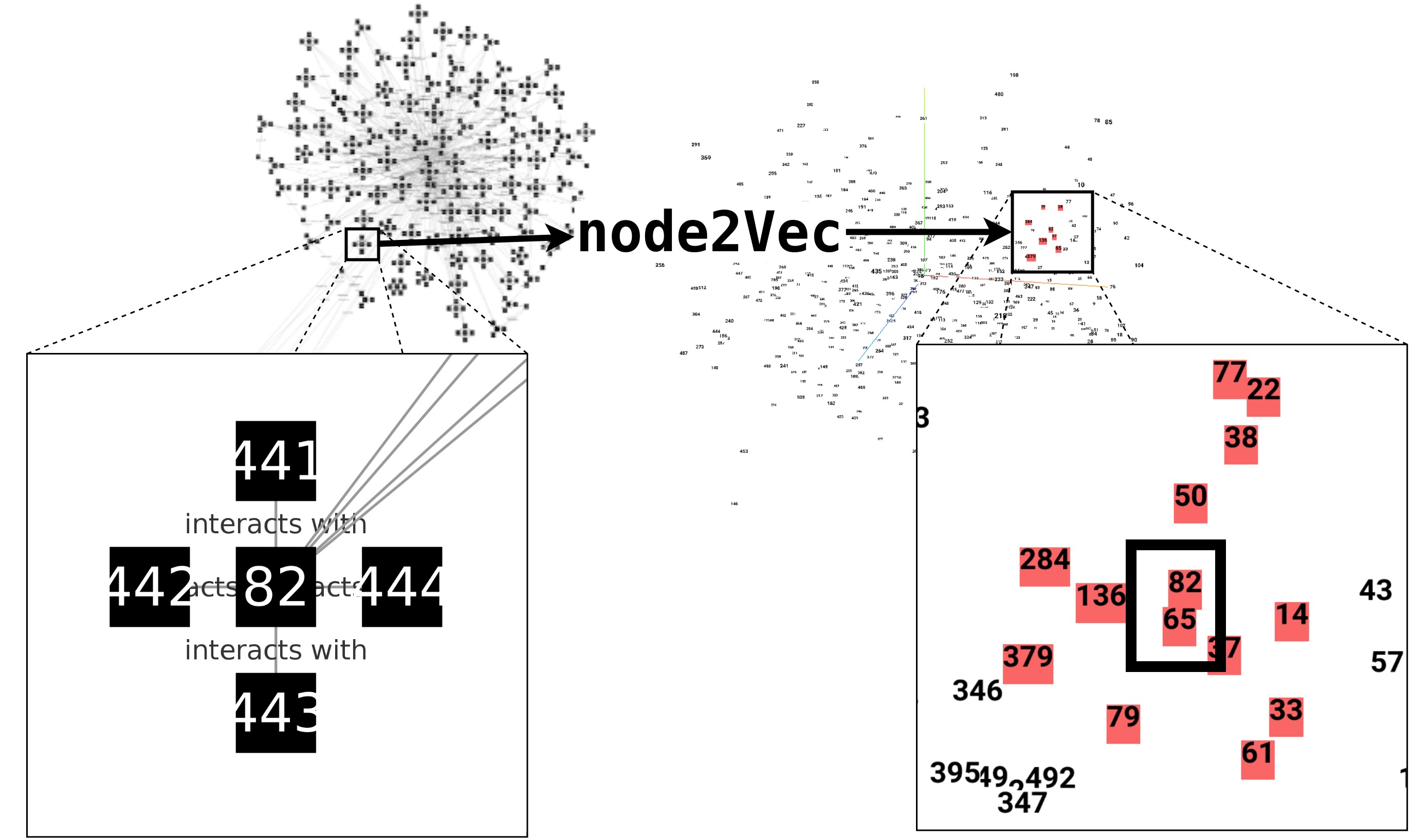}
\caption{Feature learning using node2Vec for the SDE4 building (link to a live online demo found in the text)}
\label{fig:node2vec2}
\end{figure}

\begin{figure}[h!]
\centering
\includegraphics[width=0.9\columnwidth]{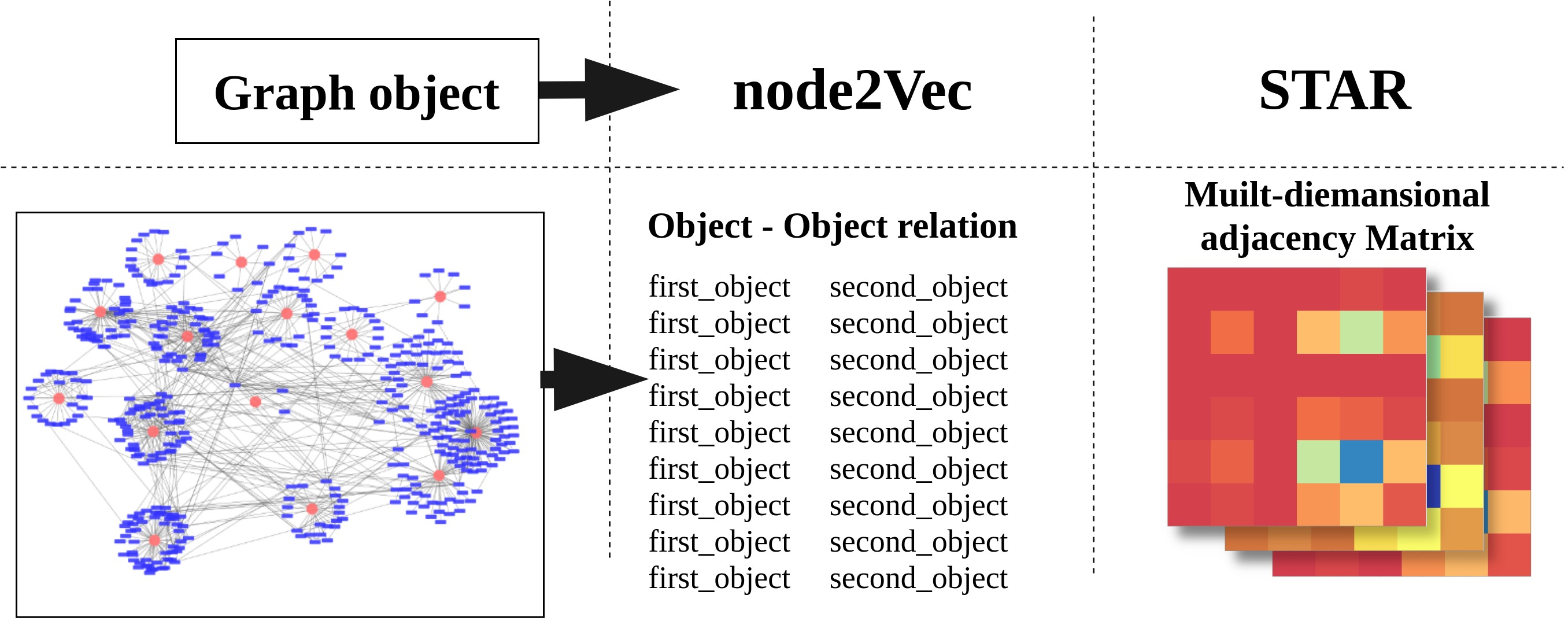}
\caption{Converting the graph object into different formats for the feature learning process (link to a higher resolution online version found in the text)}
\label{fig:graph2model}
\end{figure}

Two models of graph objects are created for both the node2Vec and STAR methods. The first model is a simple graph that represents the relation between every two objects in the form of an (\texttt{object->object}) relation as seen in Figure \ref{fig:node2vec2}. A live demo of this Figure is found online\footnote{\href{http://tiny.cc/ia0qiz}{http://tiny.cc/ia0qiz}}. To view the demo properly, deselect the \emph{Sphereize data} on the left menu. It can be noticed that nodes 82 and 65 are spatially close to each other, this means that they share the same topological structure, or in this case, \emph{IfcSpace}. A higher resolution image of Figure \ref{fig:node2vec2} can be viewed online\footnote{\href{http://tiny.cc/o874lz}{http://tiny.cc/o874lz}}. The second model consists of a multi-dimensional adjacency tensor, where the third dimension captures the temporal attributes of the graph, as seen in Figure \ref{fig:graph2model}. This tensor is then fed into a gated recurrent neural network first to capture the critical features and then to ignore useless features.


\section{Conclusion}
\label{sec:conclusion}
In this paper, we introduced and implemented a framework of an embedding graph representation of buildings into a lower-dimensional vector. This process could enhance the machine learning capabilities of the complex relationships between the spatial model, temporal data sources, and dynamic occupant behavior. This application is challenging as not many simulation tools can translate the data from these sources into a comprehensive model. Vector representation is important as a feature input in the learning process to predict, classify, or cluster relative nodes, values and/or edges attributes. We identified two possible algorithms to be used: node2Vec and STAR and show them in the context of the SDE4 building. This work is a preliminary step towards the conversion of BIM models into the vector space.





\balance

\bibliographystyle{habbrv.bst}
\bibliography{SampleReferences}

\end{document}